\def\doublespace{\baselineskip=20pt plus 3pt}
\newcommand{\sfrac}[2]{{\textstyle{#1\over#2}}}
\newcommand{\newstack}[2]{\!\!\!\begin{array}[b]{l} #1\\*[-.13cm] 
  #2\end{array}\!}
\documentstyle{article}
\begin{document}
\title {Implosion of quadrupole gravitational waves}
\author{W.B.Bonnor and M.S.Piper}
\maketitle
\setlength{\parindent}{0.5in}
\begin{abstract}
Einstein's vacuum equations are solved up to the second approximation 
for imploding quadrupole gravitational waves.  The implosion generates 
a black hole singularity irrespective of the strength of the waves.
\end{abstract}
\section{Introduction}
Abraham and Evans [1] [2] have reported numerical calculations describing
the implosion of quadrupole gravitational waves.  They find that  the
waves disperse if sufficiently weak, but form a singularity if strong 
enough. This work has implications for the formation of black holes 
and for critical phenomena in general relativity.

In this paper we describe analytical approximate solutions for incoming 
quadrupole waves which lead to a different conclusion.  We first give 
the solution of the linear approximation to Einstein's equations  
corresponding to an axially symmetric incoming quadrupole wave.  Then, 
using previously published work, we can write down the second 
approximation exactly.  We find that there is a singularity after 
the implosion of the wave, whatever its strength.

Most work on gravitational waves in the past has, naturally, dealt
with outgoing waves, and quadrupole waves have often been studied. For
reviews see [3] [4] [5]. In this paper we shall use the double-series 
approximation method [6] [7] which was set up for outgoing waves, but 
can be adapted to deal with incoming waves. We are therefore able to 
use previous work, particularly [7], with hardly more than a few changes 
of sign, to get the results of this paper.

In Section 2 we introduce the metric we use to study the imploding waves 
and in Section 3 we describe our approximation method. The linear and 
second approximations are given in Sections 4 and 5, and there is a
concluding Section 6.

\section{The metric}
We confine ourselves to the axially symmetric case, and use a metric of
the form given by Bondi [8] [9].  For the study of outgoing waves one 
can take
\begin{equation}
ds^{2}=-r^{2}(b d\theta^{2}+c\sin^{2}\theta d\phi^{2})+ e du^{2} 
+2f dr du +2rg d\theta du,
\end{equation}
where $c=b^{-1}$ and $b,e,f,g$ are functions of $r, \theta$ and $u$.  
The flat form of this is
\begin{equation}
ds^{2}=-r^{2}(d\theta^{2} + \sin^{2}\theta d\phi^{2}) + du^{2} +2drdu
\end{equation}
from which we see that $u$ is a retarded time, obtained from the $t$ of the
Minkowski metric by putting $u=t-r$.  The Schwarzschild solution in these
coordinates is
\begin{equation}
ds^{2}=-r^{2}(d\theta^{2}+\sin^{2}\theta d\phi^{2}) +(1-2mr^{-1})du^{2} 
+2dr du.
\end{equation}

In this paper we are concerned with incoming waves.  We therefore use 
an advanced time $v$ given in terms of the ordinary Minkowski $t$ by 
$v=t+r$.  The flat and Schwarzschild metrics
now become
\begin{eqnarray}
ds^{2}=-r^{2}(d\theta^{2}+\sin^{2}\theta d\phi^{2}) +dv^{2} -2drdv,\\
ds^{2}=-r^{2}(d\theta^{2}+\sin^{2}\theta d\phi^{2}) +(1-2mr^{-1})dv^{2}
-2dr dv.
\end{eqnarray}
which can be obtained from (2) and (3) by the coordinate transformation
\begin{equation}
u=-v,
\end{equation}
$r,\theta, \phi$ being unchanged.  (For a discussion of the Schwarzschild
solution in advanced and retarded time coordinates see [10]).

In what follows we shall use (6) to transform known solutions for outgoing
waves to new solutions for incoming waves.  The metric for the incoming waves
will be written
\begin{equation}
ds^{2}=-r^{2}(Bd\theta^2+C\sin^{2}\theta d\phi^{2}) +D dv^{2} +2F drdv +2rG 
d\theta dv,
\end{equation}
where $C=B^{-1}$ and $B,D,F,G$ are
functions of $r, \theta, v$.  Metrics (4) and (5) are special cases of (7).

\section{The approximation method}
In the double-series approximation method as used heretofore one seeks 
solutions of Einstein's vacuum equations for an isolated source by expanding 
the $g_{ik}$ in terms of two parameters $m$ and $a$, referring respectively 
to the source's mass and a characteristic length associated with the source:
\begin{equation}
 g_{ik}=\sum_{p=0}^{\infty}\sum_{s=0}^{\infty}m^{p}a^{s}
\stackrel{(ps)}{g}_{ik},
\end{equation}
where $\stackrel{(ps)}{g}_{ik}$ are independent of $m$ and $a$.  The flat 
background metric is obtained by putting $p=0, s=0$, the Schwarzschild 
metric in Bondi coordinates comes from $p=1, s=0$, and the linear 
approximation consists of the infinite series of terms $p=1, s>0$.  The 
latter comprise the multipole terms, and these can be prescribed arbitrarily.
By considering non-linear terms $p\geq2,s\geq1$ one can study the 
interactions of the multipole effects with themselves, or with the 
Schwarzschild mass term.  For further details, see [6]. The linear 
quadrupole terms are those with $p=1, s=2$.

For incoming waves as considered in this paper the concept of an isolated
source is not applicable because the waves are coming from infinity.
However, we can still consider the radiation in terms of multipoles, and 
study their interactions.
In this paper we are interested only in quadrupole waves, and their
interactions with themselves.  There is no $(10)$ (Schwarzschild) term,
and no other multipole terms.  We can therefore simplify the notation
by putting $\lambda=ma^{2}$, and expanding the field equations in the
single parameter $\lambda$.  Thus (8) can be written
\begin{equation}
g_{ik}=\sum_{n=0}^{\infty}\lambda^{n}\stackrel{(n)}{g}_{ik},
\end{equation}
where the $\stackrel{(n)}{g}_{ik}$ are independent of $\lambda$.  For the 
metric (7) we have
\begin{eqnarray}
-g_{22}=r^{2}B               &=&r^{2}[1+\sum_{n=1}^{\infty}\lambda^{n}
\stackrel{(n)}{B}],\\
-g_{33}=r^{2}C \sin^{2}\theta&=&r^{2}\sin^{2}\theta[1+\sum_{n=1}^{\infty}
\lambda^{n}\stackrel{(n)}{C}],\\
 g_{44}=D                    &=&1+\sum_{n=1}^{\infty}\lambda^{n}
\stackrel{(n)}{D},\\
 g_{14}=F                    &=&-1+\sum_{n=1}^{\infty}\lambda^{n}
\stackrel{(n)}{F},\\
 g_{24}=rG                   &=&r\sum_{n=1}^{\infty}\lambda^{n}
\stackrel{(n)}{G},
\end{eqnarray}
In this paper we confine ourselves to the terms with $n=1$ and $n=2$.

\section{The linear approximation}
We first substitute (7) into the vacuum
equations
\begin{equation}
R_{ik}=0
\end{equation}
and separate the linear approximation which, because of the axial symmetry, 
consists of seven second order equations
\begin{equation}
\Phi_{lm}(\stackrel{(1)}{g}_{ik})=0.
\end{equation}
The solution of these for incoming quadrupole waves can be
written down from the corresponding solution for outgoing waves given in
[6]  equations (6.6), (6.9), (6.10) and (6.11), by the use of transformation 
(6). It is
\begin{eqnarray}
\stackrel{(1)}{D}&=&-P_{2}(2r^{-1}\ddot{h}-2r^{-2}\dot{h}+r^{-3}h),\\
\stackrel{(1)}{F}&=&0,\\
\stackrel{(1)}{G}&=&(1/6)P^{\prime}_{2}(-2r^{-1}\ddot{h}-4r^{-2}\dot{h}+
3r^{-3}h),\\
\stackrel{(1)}{B}&=&(1/2) \sin^{2}\theta(r^{-1}\ddot{h}+r^{-3}h),\\
\stackrel{(1)}{C}&=&-\stackrel{(1)}{B},
\end{eqnarray}
where $P_{2}$ denotes the second Legendre polynomial, $\prime$ denotes 
differentiation with respect to $\theta$, $\lambda h(v)$ is the quadrupole 
moment, and an overhead dot means differentiation with respect to $v$.  We 
shall assume that
\begin{equation}
h=0,v<v_{1};\; \; h=0,v>v_{2}
\end{equation}
and that $h$ is a smooth function of $v$ in $v_{1}<v<v_{2}$.  Thus the 
incoming
quadrupole wave is of finite duration.  The spacetime is flat for $v<v_{1}$.
In the linear approximation (17)-(21) it is flat also for $v>v_{2}$, but this
is not so for the second approximation, as we shall see.

The functions $D,G,B$ are singular at $r=0$, and one can confirm
that $r=0$ is a curvature singularity.  It therefore appears that in the
linear approximation the incoming quadrupole waves generate a physical 
singularity
in the neighbourhood of $r=0$, whatever their strength.
However, this singularity disappears at $v=v_{2}$ because of (22).  This 
non-physical
behaviour will be corrected in the second approximation.

\section{The second approximation}
We now have to solve seven equations of the form
\begin{equation}
\Phi_{lm}(\stackrel{(2)}{g}_{ik})=\Psi_{lm}(\stackrel{(1)}{g}_{ik}),
\end{equation}
where  the left-hand side is linear in $\stackrel{(2)}{g}_{ik}$ (and their
derivatives). The right-hand side is non-linear in the 
$\stackrel{(1)}{g}_{ik}$
(and their derivatives), which are known from the linear approximation.

This problem was completely solved for outgoing quadrupole waves in [7]
and all we need to do to obtain a
solution for incoming quadrupole waves is to apply the transformation (6)
to that work.  The result of doing this is given in the Appendix.
Here we omit transient terms, i.e. those
which vanish for $v>v_{2}$, because our interest is in the formation of
permanent singularities.  The result is
\begin{eqnarray}
\stackrel{(2)}{B}&=&-\frac{1}{120r}\sin^{2}\theta(2+\sin^{2}\theta)Y,\\
\stackrel{(2)}{C}&=&-\stackrel{(2)}{B},\\
\stackrel{(2)}{D}&=&\frac{1}{15r}Y,\\
\stackrel{(2)}{F}&=&0,\\
\stackrel{(2)}{G}&=&-\frac{1}{120r}\sin\theta \cos\theta (4+3\sin^{2}\theta)Y,
\end{eqnarray}
where $s=\sin \theta, c=\cos \theta$, and
\begin{equation}
Y=-\int_{-\infty}^{v}\stackrel{...}{h}^{2}dv.
\end{equation}

$Y(v)$ in (24)-(28) vanishes for $v<v_{1}$ but is a negative constant for 
$v>v_{2}$.
This represents a permanent change in the metric.  In fact, by a
coordinate transformation similar to (9.13) of [6], {\em one can show that 
the metric for $v>v_{2}$ is, up to order $\lambda^{2}$, that of Schwarzschild
for a particle of mass $\lambda^{2}\mid Y(v_{2})\mid /30$.}  The result is 
valid whatever
the value of $\lambda$, which means that it applies whether the
waves are weak or strong.

\section{Conclusion}
We have shown that an incoming quadrupole wave produces, in the second
approximation, a singularity which persists after the wave has
ceased. {\em This result is independent of the strength of the wave.}
At this level of approximation the singularity is a black hole, since the
metric has Schwarzschild form.

The solution we have presented here is, so to speak, the mirror image of
the corresponding terms of the solution for outgoing waves.  In [6] and [7]
$Y$ (there denoted by $\stackrel{(24)}{Y}$) traces a loss of mass of an 
isolated source, represented by a
Schwarzschild particle ; this corresponds to the energy carried
away by the waves.  In the imploding case a corresponding amount of energy 
flows
inward, converging on $r=0$.

Our result, that a black hole is formed whatever the strength of the 
incoming wave,
disagrees with that of [1] and [2].  One obvious possible cause of the
disagreement is that whereas our work stops at the second approximation
the numerical calculations of Abraham and Evans include contributions
from higher approximations.  However, this explanation seems implausible
for the following reason. The disagreement occurs when the incoming
wave is weak, i.e. for {\em small} values of $\lambda$; but for these values
one would expect convergence to be quicker, and the higher
powers in the expansion to be less important.\vspace{0.3in}

We gratefully acknowledge an interesting correspondence with Dr. John M 
Stewart.\\
\vspace{0.3in}\\
{\sc\large Appendix}\\
Here we give the complete solution to the second approximation summarised
in Section 5.  It is obtained by using the transformation (6) on the
corresponding solution of the second approximation for the outgoing wave
problem, namely (3.7)-(3.11) of [7].  The only matter which is not quite
straightforward is the relation of  $\stackrel{24}{Y}$ of [7]
and $Y$ in Section 5.  The essential point here is that the former is
defined by
\begin{equation}
\frac{d\stackrel{(24)}{Y}}{du}=(h^{\prime\prime\prime})^{2}
\end{equation}
so on applying the transformation we have
\begin{equation}
-\frac{dY}{dv}=\stackrel{...}{h}^{2}.
\end{equation}
We are now at liberty to integrate this equation:
\begin{equation}
Y=-\int_{-\infty}^{v}\stackrel{...}{h}^{2}dv
\end{equation}
the lower limit being chosen because spacetime is assumed to be empty and 
static
at $v=-\infty$. Thus $\stackrel{(24)}{Y}$ is replaced by $Y$ defined in (29).

The complete second approximation is
\doublespace
\begin{eqnarray}
\stackrel{(2)}{D}&=&[(\sfrac{3}{5}-3s^{2}+\sfrac{21}{8}s^{4})
\dot{h}h^{\mbox{{\tiny IV}}}+(\sfrac{12}{5}-9s^{2}+\sfrac{27}{4}s^{4})
\ddot{h}\newstack{...}{h}+\sfrac{1}{15}Y]r^{-1}\nonumber \\ 
& &+[(-2+10s^{2}-\sfrac{35}{4}s^{4})\dot{h}\newstack{...}{h}+(-2+9s^{2}
-\sfrac{15}{2}s^{4})\ddot{h}^{2}]r^{-2}\nonumber \\
& &+(\sfrac{9}{2}-\sfrac{47}{2}s^{2}+\sfrac{331}{16}s^{4})\dot{h}
\ddot{h}r^{-3}+[(\sfrac{3}{2}s^{2}-\sfrac{27}{16}s^{4})h\ddot{h}\nonumber \\
& &+(-\sfrac{15}{4}+\sfrac{71}{4}s^{2}-\sfrac{493}{32}s^{4})\dot{h}^{2}]
r^{-4}+(\sfrac{3}{2}-6s^{2}+\sfrac{81}{16}s^{4})\dot{h}hr^{-5}\nonumber \\
& &+(\sfrac{1}{2}-3s^{2}+\sfrac{21}{8}s^{4})h^{2}r^{-6}\\
\stackrel{(2)}{F}&=&s^{4}[\sfrac{1}{32}\ddot{h}^{2}r^{-2}+\sfrac{3}{32}h
\ddot{h}r^{-4}+\sfrac{3}{32}h^{2}r^{-6}]\\
\stackrel{(2)}{G}&=&[(-\sfrac{3}{10}cs+\sfrac{21}{40}cs^{3})\dot{h}
h^{\mbox{{\tiny IV}}}+(-\sfrac{6}{5}cs+\sfrac{27}{20}cs^{3})\ddot{h}
\newstack{...}{h}-(\sfrac{1}{30}cs+\sfrac{1}{40}cs^{3})Y]r^{-1}\nonumber \\
& &+[(-2cs+\sfrac{7}{2}cs^{3})\dot{h}\newstack{...}{h}+(-2cs+
\sfrac{23}{8}cs^{3})\ddot{h}^{2}]r^{-2}\nonumber \\
& &+(\sfrac{27}{4}cs-\sfrac{185}{16}cs^{3})\dot{h}\ddot{h}r^{-3}+
[\sfrac{1}{8}cs^{3}h\ddot{h}+(-\sfrac{15}{2}cs+\sfrac{105}{8}cs^{3})
\dot{h}^{2}]r^{-4}\nonumber \\
& &+(\sfrac{15}{4}cs-\sfrac{107}{16}cs^{3})h\dot{h}r^{-5}+
(\sfrac{3}{2}cs-\sfrac{21}{8}cs^{3})h^{2}r^{-6}\\
\stackrel{(2)}{B}&=&[(-\sfrac{3}{20}s^{2}+\sfrac{7}{40}s^{4})
\dot{h}h^{\mbox{{\tiny IV}}}+(-\sfrac{3}{5}s^{2}+\sfrac{9}{20}s^{4})
\ddot{h}\newstack{...}{h}+(-\sfrac{1}{60}s^{2}-\sfrac{1}{120}s^{4})Y]r^{-1}
\nonumber \\
& &+\sfrac{1}{8}s^{4}\ddot{h}^{2}r^{-2}+(-\sfrac{9}{4}s^{2}+
\sfrac{21}{8}s^{4})\dot{h}\ddot{h}r^{-3}+[\sfrac{1}{4}s^{4}h\ddot{h}+
(\sfrac{75}{16}s^{2}-\sfrac{175}{32}s^{4})\dot{h}^{2}]r^{-4}\nonumber \\
& &+(-\sfrac{27}{8}s^{2}+\sfrac{69}{16}s^{4})\dot{h}hr^{-5}+
(-\sfrac{7}{4}s^{2}+\sfrac{15}{8}s^{4})h^{2}r^{-6}\\
\stackrel{(2)}{C}&=&[(\sfrac{3}{20}s^{2}-\sfrac{7}{40}s^4)
\dot{h}h^{\mbox{\tiny IV}}+(\sfrac{3}{5}s^2-\sfrac{9}{20}s^4)
\ddot{h}\newstack{...}{h}+(\sfrac{1}{60}s^2+\sfrac{1}{120}s^4)Y]
r^{-1}\nonumber \\
& &+\sfrac{1}{8}s^4\ddot{h}^2r^{-2}+(\sfrac{9}{4}s^2-\sfrac{21}{8}s^4)
\dot{h}\ddot{h}r^{-3}+[\sfrac{1}{4}s^4h\ddot{h}+(-\sfrac{75}{16}s^2+
\sfrac{175}{32}s^4)\dot{h}^2]r^{-4}\nonumber \\
& &+(\sfrac{27}{8}s^2-\sfrac{69}{16}s^4)\dot{h}hr^{-5}+(\sfrac{7}{4}s^2-
\sfrac{13}{8}s^4)h^2r^{-6}
\end{eqnarray}
\baselineskip=12pt
where $s\equiv\sin\theta$ and $c\equiv\cos\theta$.\vspace{0.5in}

{\sc\large References}\\
{[1]}Abraham A M and Evans C R 1992 {\em Phys. Rev.} D {\bf 46} R4117\\
{[2]}Abraham A M and Evans C R 1993 {\em Phys. Rev. Lett.} {\bf 70} 2980\\
{[3]}Thorne K S 1980 {\em Rev. Mod. Phys.} {\bf 52} 285\\
{[4]}Damour T 1986 in{\em Gravitation in Astrophysics} ed. B Carter and
J M Hartle (Plenum)\\
{[5]}Blanchet L 1996 Lecture notes: gr-qc/9607025\\
{[6]}Bonnor W B and Rotenberg M A 1966 {\em Proc. Roy. Soc.} A {\bf 289} 247\\
{[7]}Hunter A J and Rotenberg M A 1969 {\em J. Phys. Soc.} A {\bf 2} 34\\
{[8]}Bondi H 1960 {\em Nature} {\bf 186} 535\\
{[9]}Bondi H, van der Burg and Metzner A W K  1962 {\em Proc. Roy. Soc.} A 
{\bf 269} 21\\
{[10]}Misner C W, Thorne K S and Wheeler J 1970 {\em Gravitation} (Freeman) 
pages 828-829

\end{document}